\documentclass{article}

\usepackage{amssymb,amsmath,xcolor,graphicx,xspace,colortbl,rotating} 
\usepackage{amsmath}  
\usepackage{tabulary}  
\graphicspath{{deutschoctober_graphics/}{deutschoctober_tcache/}{deutschoctober_gcache/}}
\DeclareGraphicsExtensions{.pdf,.eps,.ps,.png,.jpg,.jpeg}

\begin{document}
\title{Back to the seminal Deutsch algorithm
}

\author{Giuseppe Castagnoli
\\
}
\maketitle

\begin{abstract} A bare description of the seminal
quantum algorithm devised by Deutsch could mean more than an introduction to quantum computing. It
could contribute to opening the field to interdisciplinary research.
\end{abstract}

\section{Motivation
}
The usual introductions to quantum computation are necessarily burdened by the
mathematical gear required for a comprehensive description of it, like quantum computational
networks or the quantum Turing machine. This inevitably increases the cost of accessing the subject.
We think that providing a bare description of the seminal quantum algorithm devised by Deutsch
$\left [1\right ]$
is the best way of both introducing the subject and opening it to interdisciplinary
research. It is reasonable to think that knowledge of this elementary quantum algorithm, the
prototype of all the subsequent quantum algorithms, is sufficient to investigate the foundations of
quantum computation.

 Section 2 describes the problem solved by Deutsch algorithm, Section 3 the quantum
algorithm itself, and Section 4 is a discussion of the fundamental questions raised by its quantum
computational speedup.

\section{ The problem
}
The problem solved by Deutsch algorithm is as follows. We have the set of
functions of table (\ref{table}).

\begin{equation}
\begin{tabular}[c]{lllll}
$a$
&
$f_{00}\left (a\right )$
&
$f_{01}\left (a\right )$
&
$f_{10}\left (a\right )$
&
$f_{11}\left (a\right )$
\\
$0$
&
$0$
&
$0$
&
$1$
&
$1$
\\
$1$
&
$0$
&
$1$
&
$0$
&
$1$
\end{tabular}
\label{table}
\end{equation}

$f_{00}\left (a\right )$
and
$f_{11}\left (a\right )$
are constant functions,
$f_{01}\left (a\right )$
and
$f_{10}\left (a\right )$
, with an even number of zeros and ones, are \textit{balanced} functions.

 Bob, the problem setter, chooses one of these functions and gives Alice, the
problem solver, a black box\protect\footnote{ The box is called black because its inside (i. e. the function chosen by
Bob) must be hidden to Alice.
} that given a value of the argument
$a$
produces the corresponding value of the function. Alice knows the set of functions but
ignores Bob's choice. She is to find whether the function chosen by Bob is constant or balanced by
computing the value of the function for suitable values of the argument (namely, by performing
\textit{function evaluations}).

Logically, and in the case of classical computation, Alice has to perform two
function evaluations, for
$a =0$
and
$a =1$
. In the quantum case, just one is enough, for a quantum superposition of the two
possible values of the argument. There is the so called \textit{quantum computational
speedup}.

\section{ Deutsch algorithm
}

An algorithm is a prescribed sequence of
arithmetical operations performed on a set of registers that contain the relevant numbers. Presently
we need: (i) a two bit register
$B$
that contains the \textit{problem setting}, namely the suffix
$b$
of the function chosen by Bob, (ii) a one bit register
$A$
that contains the argument
$a$
for which the function should be computed by the black box, and (iii) a one bit
register
$V$
meant to contain the result of the computation, modulo
$2$
added to the former register's content for logical reversibility. Register
$B$
is absent in the original algorithm. We have introduced it in view of discussing the
reason for the computational speedup.

In the quantum case, each register is
characterized by its quantum state. For example, in the quantum state
$\left \vert 00\right \rangle _{B}$
, the value of
$b$
contained in register
$B$
is
$00$
. The vector
$\left \vert 00\right \rangle _{B}$
is called a \textit{basis vector} of register
$B$
; the other three being
$\left \vert 01\right \rangle _{B} ,\left \vert 10\right \rangle _{B} ,$
and
$\left \vert 11\right \rangle _{B}$
. The content of register
$B$
can be acquired by measuring the observable
$\hat{B}$
of eigenstates
$\left \vert 00\right \rangle _{B} ,\left \vert 01\right \rangle _{B} ,$
etc. and eigenvalues respectively
$00 ,01 ,$
etc. Similarly, the basis vectors of register
$A$
are
$\left \vert 0\right \rangle _{A}$
and
$\left \vert 1\right \rangle _{A}$
, those of register
$V$
are
$\left \vert 0\right \rangle _{V}$
and
$\left \vert 1\right \rangle _{V}$
.

Deutsch algorithm goes as follows.

In the assumption that, say,
$b =01$
, the initial state of the three registers is:

\begin{equation}\left \vert \psi \right \rangle  =\frac{1}{\sqrt{2}}\left \vert 01\right \rangle _{B}\left \vert 0\right \rangle _{A}\left (\left \vert 0\right \rangle _{V} -\left \vert 1\right \rangle _{V}\right ) \label{inb}
\end{equation}

 In view of what will follow, the state of register
$B$
should be seen as the random outcome of the initial measurement of
$\hat{B}$
on the part of Bob,             in a state where the problem setting is completely
undetermined. It is simpler to think that Bob selects a problem setting at random; he could change
it unitarily into a desired setting but this is irrelevant to present ends. The initial state of
register
$A$
can be any basis vector of it, we have chosen
$\left \vert 0\right \rangle _{A}$
. The purpose of the particular initial state of register
$V$
will soon become clear.

State (\ref{inb}) is thus the input state of
the quantum algorithm prepared by Bob. The first operation performed by Alice is
the application of the Hadamard transform
$H_{A}$
to register
$A$
. This is a unitary transformation defined as follows:
$H_{A}\left \vert 0\right \rangle _{A} =\frac{1}{\sqrt{2}}\left (\left \vert 0\right \rangle _{A} +\left \vert 1\right \rangle _{A}\right )$
and
$H_{A}\left \vert 1\right \rangle _{A} =\frac{1}{\sqrt{2}}\left (\left \vert 0\right \rangle _{A} -\left \vert 1\right \rangle _{A}\right )$
. Thus
$H_{A}$
sends state (\ref{inb}) into:

\begin{equation}H_{A}\left \vert \psi \right \rangle  =\frac{1}{2}\left \vert 01\right \rangle _{B}\left (\left \vert 0\right \rangle _{A} +\left \vert 1\right \rangle _{A}\right )\left (\left \vert 0\right \rangle _{V} -\left \vert 1\right \rangle _{V}\right ) \label{hadamard}
\end{equation}

The second operation is function evaluation
$H_{f}$
. It is performed \textit{in quantum parallelism} on each and every element of
the quantum superposition (\ref{hadamard}), as follows.

For example, the element
$\left \vert \psi _{01 ,1 ,0}\right \rangle  =$
$\left \vert 01\right \rangle _{B}\left \vert 1\right \rangle _{A}\left \vert 0\right \rangle _{V}$
tells us that the function chosen by Bob is
$f_{01}\left (a\right )$
(suffix
$01$
in register
$B$
), that the argument of the function to be computed by the black box is
$1$
(see the content of register
$A$
), and that the content of register
$V$
is
$0$
.

The function evaluation transformation is therefore
$H_{f}\left \vert \psi _{01 ,1 ,0}\right \rangle  =$
$\left \vert 01\right \rangle _{B}\left \vert 1\right \rangle _{A}\left \vert 1\right \rangle _{V}$
. In fact, by definition, the basis vectors of registers
$B$
and
$A$
go unaltered through function evaluation; the computation of
$f_{01}\left (1\right )$
yields
$1$
(table \ref{table}), which -- modulo
$2$
added to the former content of register
$V$
-- yields
$1$
-- so that
$\left \vert 0\right \rangle _{V}$
goes into
$\left \vert 1\right \rangle _{V}$
.

Let us also note that function evaluation is logically reversible, thus unitary,
since the output of the transformation keeps the memory of the input.

Similarly, the element
$ -\left \vert 01\right \rangle _{B}\left \vert 1\right \rangle _{A}\left \vert 1\right \rangle _{V}$
goes into
$ -\left \vert 01\right \rangle _{B}\left \vert 1\right \rangle _{A}\left \vert 0\right \rangle _{V}$
, etc. Note that, as a consequence and more in general,
$\frac{1}{\sqrt{2}}\left \vert b\right \rangle _{B}\left \vert a\right \rangle _{A}(\left \vert 0\right \rangle _{A} -\left \vert 1\rangle _{A}\right )$
goes into itself when
$f_{b}\left (a\right ) =0$
, into
$\frac{1}{\sqrt{2}}\left \vert b\right \rangle _{B}\left \vert a\right \rangle _{A}\left (\left \vert 1\right \rangle _{V} -\left \vert 0\right \rangle _{V}\right ) = -$
$\frac{1}{\sqrt{2}}\left \vert b\right \rangle _{B}\left \vert a\right \rangle _{A}\left (\left \vert 0\right \rangle _{V} -\left \vert 1\right \rangle _{V}\right )$
when
$f_{b}\left (a\right ) =1$
. In the overall, function evaluation sends state (\ref{hadamard}) into:

\begin{equation}H_{f}H_{A}\left \vert \psi \right \rangle  =\frac{1}{2}\left \vert 01\right \rangle _{B}\left (\left \vert 0\right \rangle _{A} -\left \vert 1\right \rangle _{A}\right )(\left \vert 0\right \rangle _{V} -\left \vert 1\right \rangle _{V}) \label{sel}
\end{equation}

 Applying another time the Hadamard transform to register
$A$
sends state
(\ref{sel}) into:

\begin{equation}H_{A}H_{f}H_{A}\left \vert \psi \right \rangle  =\frac{1}{\sqrt{2}}\left \vert 01\right \rangle _{B}\left \vert 1\right \rangle _{A}(\left \vert 0\right \rangle _{V} -\left \vert 1\right \rangle _{V}) \label{output}
\end{equation}

We will see in a moment that the state of
register
$A$
, namely
$\left \vert 1\right \rangle _{A}$
, encodes the solution of the problem, the fact that the function is balanced. Alice
eventually acquires the solution by measuring the observable
$\hat{A}$
-- of eigenstates
$\left \vert 0\right \rangle _{A}$
,
$\left \vert 1\right \rangle _{A}$
and eigenvalues respectively
$0$
,
$1$
-- in state (\ref{output}) (note that
$\hat{B}$
and
$\hat{A}$
commute). She acquires the eigenvalue
$1$
, which tells her that the function is balanced.

In view of the Discussion, we note that the
quantum state remains unaltered throughout the measurement of
$\hat{A}$
-- the state of register
$A$
immediately before measurement is always an eigenstate of
$\hat{A}$
. There is thus a unitary transformation between the outcome of the initial measurement of
$\hat{B}$
, namely state (\ref{inb}), and the outcome of the final measurement of
$\hat{A}$
. The process between them is physically reversible since no information is destroyed along
it.

 We also note that the reduced density operator of register
$B$
remains unaltered through
$H_{A}H_{f}H_{A}$
: its basis vectors go unaltered through
$H_{f}$
and the two
$H_{A}$
only apply to the basis vectors of register
$A$
.

That the eigenvalue
$1$
of
$\hat{A}$
tells \textit{balanced} and
$0$
\textit{constant} can be seen by writing Deutsch algorithm for all the possible
choices of the value of
$b$
. We do this by performing Deutsch algorithm for a quantum superposition of all the basis
vectors of register
$B$
. The initial state becomes:

\begin{equation}\left \vert \Psi \right \rangle  =\frac{1}{2\sqrt{2}}\left (\left \vert 00\right \rangle _{B} +\left \vert 01\right \rangle _{B} +\left \vert 10\right \rangle _{B} +\left \vert 11\right \rangle _{B}\right )\left \vert 0\right \rangle _{A}\left (\left \vert 0\right \rangle _{V} -\left \vert 1\right \rangle _{V}\right )
\end{equation}

The successive states:

\begin{equation}H_{A}\left \vert \Psi \right \rangle  =\frac{1}{4}\left (\left \vert 00\right \rangle _{B} +\left \vert 01\right \rangle _{B} +\left \vert 10\right \rangle _{B} +\left \vert 11\right \rangle _{B}\right )\left (\left \vert 0\right \rangle _{A} +\left \vert 1\right \rangle _{A}\right )\left (\left \vert 0\right \rangle _{V} -\left \vert 1\right \rangle _{V}\right )
\end{equation}

\begin{equation}H_{f}H_{A}\left \vert \Psi \right \rangle  =\frac{1}{4}[\left (\left \vert 00\right \rangle _{B} -\left \vert 11\right \rangle _{B}\right )\left (\left \vert 0\right \rangle _{A} +\left \vert 1\right \rangle _{A}\right ) +(\left \vert 01\right \rangle _{B} -\left \vert 10\right \rangle _{B})\left (\left \vert 0\right \rangle _{A} -\left \vert 1\right \rangle _{A}\right )](\left \vert 0\right \rangle _{V} -\left \vert 1\right \rangle _{V}) , \label{parallel}
\end{equation}

\begin{equation}H_{A}H_{f}H_{A}\left \vert \Psi \right \rangle  =\frac{1}{2\sqrt{2}}[\left (\left \vert 00\right \rangle _{B} -\left \vert 11\right \rangle _{B}\right )\left \vert 0\right \rangle _{A} +(\left \vert 01\right \rangle _{B} -\left \vert 10\right \rangle _{B})\left \vert 1\right \rangle _{A}](\left \vert 0\right \rangle _{A} -\left \vert 1\rangle _{A}\right )
\label{out}
\end{equation}

We can see that the state of registers
$B$
and
$A$
in (\ref{out}) is a quantum superposition of four tensor
products, each the product of a choice of the function computed by the black box (the value of
$b$
in register
$B$
) and the corresponding solution of the problem (the number in register
$A$
:
$0$
if the function is constant,
$1$
if it is balanced).

 The important thing is of course the fact that the solution is reached with just
one function evaluation. This is a revolutionary result, logically and in classical physics two
successive function evaluations are required.

\section{ Discussion
}

 The discovery of the first quantum speedup raised a natural question. The
speedup is of course in the mathematics of the quantum algorithm. However, conceptually, what is the
reason for it?

The first thing to say is that today, thirty two years after the publication
of
$\left [1\right ]$
, there is no accepted answer to the above question.

 A partial answer is \textit{quantum parallel
computation}. This is the fact that also Deutsch algorithm performs the two function
evaluations required in the classical case, but it does that simultaneously for a quantum
superposition of the two function arguments.

 Although it exactly explains the speedup of Deutsch
algorithm, in the general case of quantum oracle computing\protect\footnote{ An oracle problem is a generalization of Deutsch's problem. Given a set
of functions known to both Bob and Alice, Alice should find a characteristic of the function chosen
by Bob (e. g. the function period) by performing function evaluations. Most quantum algorithms solve
oracle problems.
} quantum parallel computation does not account for the number of
function evaluations required to solve the problem. The speedup is a quantitative feature and its
explanation should be quantitative in character.

It should be noted that all the quantum algorithms discovered so far have been
found by means of ingenuity. Of course their speed up is always in their mathematics, but the
reasons for it are very different from algorithm to algorithm. We report an authoritative conclusion
about the possibility of unifying the explanation of the speedup. Quoting from
$\left [2\right ]\text{:}$
\textit{The speedup appears to always depend on
the exact nature
of the problem while the reason for it varies from problem to problem}.
This was written in 2001, but in our judgment the situation has not changed since then. In
mainstream literature, there is no unifying, quantitative explanation of the speedup, neither a
fundamental physical explanation of it.

 Of course one might also think that there is none. The speedup could be an
epiphenomenon emerging at a certain complexity level, non reducible to some fundamental quantum
feature. We believe that, after thirty two years without an explanation, this way of thinking is on
the increase.

Important advances have been made instead on the quantum computer science
front: identifying quantum complexity classes and relating them to the classical ones. There is an
important body of literature on this, we provide
$\left [3\right ]$
as an example. As things are now, these studies
concern the mathematics of quantum algorithms and are not related to their physical interpretation.

 The evolutionary approach
$\left [4 \div 6\right ]$
stands on its own. In
$\left [6\right ]$
, we provide a fundamental and quantitative explanation of the speedup that, given
any oracle problem, allows to compute the number of function evaluations required to solve it in an
optimal quantum way. The other side of the coin is that the explanation in question is very
unconventional.

We outline it. It comes out from a radical application of the trademark
of quantum computation pointed out by Deutsch in his seminal 1985 paper
$\left [1\right ]$
. The all with quantum computation would be representing abstract computational
notions physically. The explanation comes out by physically representing, besides the
computation, the notion of black box, namely the fact that the problem setting (here the random
outcome of the initial measurement) is hidden to the
problem solver. We represent this concealment by postponing the projection of the quantum state due
to the initial Bob's measurement at the end of the unitary part of Alice's problem solving action --
an always legitimate operation. As a consequence, the input state of the quantum algorithm to Alice
becomes one of complete ignorance of the problem setting selected by Bob (we are under the
assumption that the state before the measurement of
$\hat{B}$
is one of complete indetermination of the problem setting). By the way, the fact that
the quantum state depends on the observer -- whether it is Bob or Alice -- is foreseen by relational
quantum mechanics
$\left [7\right ]$
.

For reasons of time-symmetry
$\left [8 \div 11\right ]$
applying to the reversible process comprised between the initial and final
measurement outcomes, a part of the random outcome of the initial measurement corresponding to half
solution (half of the information specifying it when it is an unstructured bit string) should be
selected back in time by the final measurement. As a consequence, the input state to Alice, of
complete ignorance of the outcome of the initial measurement, is projected on one where she knows
that part of it (and thus the corresponding half solution).  It turns out that quantumly it is
possible to shield an observer form the information coming to her from the past measurement,
not from that coming to her from the future measurement.

By the way, such an advanced knowledge of
half solution vanishes in the ordinary representation of the quantum process -- that with respect to
Bob -- where no observer is shielded from any measurement outcome. The information coming to the
observer from the final measurement (a part of the problem setting) is completely masked by that
coming to him from the initial measurement (the entire problem setting).

As a consequence of the above, an optimal quantum algorithm would require the
number of function evaluations required by a classical algorithm endowed with the advanced knowledge
of half of the solution of the problem. This would be a quantitative explanation of the speedup
coming out from a fundamental time-symmetry.

\section*{References
}
$\left [1\right ]$
Deutsch D. Quantum theory, the Church Turing principle and the
universal quantum computer. Proc. Roy. Soc. A 400, 97-117(1985)

$\left [2\right ]$
Henderson L, Vedral V. Classical, quantum and total correlations.
Journal of Physics A 34, 6899- 6709 (2001)

$\left [3\right ]$
Aaronson S, Ambainis A. Forrelation: a Problem that Optimally Separates Quantum from
Classical Computing. arXiv:1411.5729 [quant.ph] (2014)

$\left [4\right ]$
Castagnoli G, Finkelstein D R. Theory of the quantum speedup. Proc.
Roy. Soc. A 1799, 457, 1799-1807 (2001)

$\left [5\right ]$
Castagnoli G. The quantum correlation between the selection of the problem and that of
the solution sheds light on the mechanism of the quantum speed up. Phys. Rev. A 82,
052334-052342 (2010)

$\left [6\right ]$
Castagnoli G. Completing the physical representation of quantum algorithms provides a
quantitative explanation of their computational speedup. https://arxiv.org/pdf/1705.02657.pdf
(2017)

$\left [7\right ]$
Rovelli C. Relational Quantum Mechanics. Int. J. Theor. Phys. 35, 637-658 (1996)

$\left [8\right ]$
Aharonov Y, Bergman PG, Lebowitz JL. Time Symmetry in the Quantum Process of
Measurement. Phys. Rev. B 134,\textbf{\ }1410-1416 (1964)

$\left [9\right ]$
Dolev S, Elitzur AC. Non-sequential behavior of the wave function.
arXiv:quant-ph/0102109 v1 (2001)

$\left [10\right ]$
Aharonov Y, Vaidman L. The Two-State Vector Formalism: An
Updated Review. Lect. Notes Phys. 734, 399--447 (2008)

$\left [11\right ]$
Aharonov Y, Cohen E, Elitzur AC. Can a future choice affect a past measurement outcome?
Ann. Phys. 355, 258-268 (2015)

\end{document}